\documentclass[prd,nofootinbib,amsfonts,notitlepage]{revtex4-1}
\usepackage[utf8]{inputenc}
\usepackage[T1]{fontenc}
\usepackage{hyperref}
\usepackage{graphicx,bm,amsmath,color}
\usepackage{bbold}
\usepackage{wasysym}
\usepackage{verbatim}
\newcommand{\be}{\begin{equation}}
\newcommand{\ee}{\end{equation}}
\newcommand{\beq}{\begin{eqnarray}}
\newcommand{\eeq}{\end{eqnarray}}
\newcommand{\ba}{\begin{align}}
\newcommand{\ea}{\end{align}}

\begin{document}

\title{Does a deformation of special relativity imply energy dependent photon time delays?}
\author{J.M. Carmona}
\affiliation{Departamento de F\'{\i}sica Te\'orica,
Universidad de Zaragoza, Zaragoza 50009, Spain}
\author{J.L. Cort\'es}
\affiliation{Departamento de F\'{\i}sica Te\'orica,
Universidad de Zaragoza, Zaragoza 50009, Spain}
\author{J.J. Relancio}
\email{jcarmona@unizar.es, cortes@unizar.es, relancio@unizar.es}
\affiliation{Departamento de F\'{\i}sica Te\'orica,
Universidad de Zaragoza, Zaragoza 50009, Spain}

\begin{abstract}
Theoretical arguments in favor of energy dependent photon time delays from a modification of special relativity (SR) have met with recent gamma ray observations that put severe constraints on the scale of such deviations. We review the case of the generality of this theoretical prediction in the case of a deformation of SR and find that, at least in the simple model based on the analysis of photon worldlines which is commonly considered, there are many scenarios compatible with a relativity principle which do not contain a photon time delay. This will be the situation for any modified dispersion relation which reduces to $E=|\vec{p}|$ for photons, independently of the quantum structure of spacetime. This fact opens up the possibility of a phenomenologically consistent relativistic generalization of SR with a new mass scale many orders of magnitude below the Planck mass.
\end{abstract}

%Keywords: Deformed or deformed special relativity (DSR); gamma ray bursts and gamma ray propagation; quantum gravity phenomenology; noncommutativity; kappa-Minkowski; Lorentz symmetry  

\maketitle

\section{Introduction}

Special relativity (SR) postulates Lorentz invariance as an exact symmetry of Nature. It is at the base of our quantum field theories of the fundamental interactions, and has surpassed all experimental tests up to date (\cite{Kostelecky:2008ts,*Long:2014swa,*Kostelecky:2016pyx,*Kostelecky:2016kkn}; see also the papers in~\cite{LectNotes702}).
A quantum gravity theory, however, is expected quite generally to modify this symmetry. A heuristic reasoning is that Lorentz invariance is a symmetry of spacetime, and the classical picture of a continuum spacetime must break down somehow at the Planck scale, where quantum effects of gravity (in the form of creation and evaporation of virtual black holes, for example~\cite{Kallosh:1995hi}) should take place. Many approaches to quantum gravity incorporate a departure of Lorentz invariance in a form or another (for a review, see Ref.~\cite{AmelinoCamelia:2008qg}), and the experimental confirmation or refutation of this hypothesis would be of great importance to constrain possible theoretical frameworks for a quantum theory of gravity.

From this point of view, the symmetries of SR would constitute a good long-distance, or low-energy, approximation that could be modified at a certain high-energy scale $\Lambda$. The naive expectation would be that $\Lambda$ is around the Planck mass, $m_P\approx 1.2\times 10^{19}$\,GeV/$c^2$, which could lead to think that there are no ways to observe effects from such a modification of SR. However, over the past few years it has been realized that there are
astrophysical observations that could be sensitive to such deviations~\cite{Mattingly:2005re}. For example, thresholds of reactions can be significantly changed by modifications of SR, since their magnitude can be comparable with the contribution of masses, which, in spite of being very small in high-energy processes, are the quantities that define these thresholds (the well-known GZK cutoff~\cite{Greisen:1966jv,*Zatsepin:1966jv,*Kifune:1999ex,*Aloisio:2000cm} is an example of this).

However, the discussion of these sensitivities is very different depending on whether the modification of SR consists in a \textit{violation} or a \textit{deformation} of SR. In the first case, there is not an equivalence of inertial frames and the threshold of a reaction can get corrections of the order of $E^3/m^2\Lambda$, where $E$ is one of the energies involved in the process in our (Earth-based) laboratory frame, and $m$ is a mass that controls the corresponding threshold in SR. Moreover, reactions which are forbidden by SR can be allowed in the Lorentz-violating theory at high enough energies. In contrast, theories with a deformation of special relativity (DSR) contain a relativity principle. This means that DSR theories cannot produce a threshold for particle decays at a certain energy of the decaying particle, since the value of this energy would not be relativistically invariant. In the same way, the existence of a relativity principle implies cancellations between the effects of modified particle dispersion relations and modified conservation laws that evade many of the constraints of the Lorentz-violating case~\cite{Carmona:2010ze,*Carmona:2014aba}.

Another difference is that the privileged frame of Lorentz invariant violations may produce sidereal variations which are looked for in terrestrial experiments, while theories with a relativity principle do not exhibit such effects. Therefore, the identification of anomalies in the case of a deformed symmetry has to rely on the existence of amplification mechanisms. For example, a (small) energy dependence of the speed of photons could produce an observable time delay in the arrival of two photons of different energies emitted simultaneously from a sufficiently far away source. In this case the smallness of the modification of SR could be compensated by the amplification given by the long time of flight. Similarly, a possible birefringence (different speeds for different photon polarizations) could erase linear polarization from distant astrophysical objects. However, while photon birefringence is a prediction of effective field theories in certain Lorentz-violating scenarios, DSR theories do not necessarily include such an effect. This leaves time of flight observations as the only window to modifications of SR that are compatible with a relativity principle~\cite{Mattingly:2005re}. It is therefore of great relevance that recent measurements of the time structure of arrival of photons from gamma ray bursts by the Fermi or the MAGIC telescopes have reached enough precision to put strong bounds on first order corrections in the photon energy over the Planck mass~\cite{Albert:2007qk,*Martinez:2008ki,*Ackermann:2009aa,*HESS:2011aa,*Nemiroff:2011fk,*Vasileiou:2013vra,*Vasileiou:2015wja}.

The previous experimental results suggest that there are no leading order planckian corrections in the dispersion relation of photons, at least if interpreted in the context of a Lorentz-violating scenario. However, the situation is again more subtle in the case of a deformation of SR.  
The discussion of energy-dependent photon time delays in DSR does not only involve the possible modification of the dispersion relation, but also the implementation of nontrivial translations between observers that are local to the emission and detection of the photons. Nontrivial translations are a necessary ingredient in a modification of SR that contains deformed Lorentz transformations in order to avoid inconsistencies with tests of locality, as the discussion in Ref.~\cite{AmelinoCamelia:2010qv,*Hossenfelder:2010tm} showed. This ingredient is quite natural in a theory which deforms the Poincaré algebra of SR, and in fact it appears in the so-called relative locality framework~\cite{AmelinoCamelia:2011bm}, which is a proposal for the spacetime structure of DSR theories. In this proposal, it is the curvature of momentum space (which stems from the modified composition laws for momenta, unavoidable in DSR theories) that produce nonlocal effects in observers translated with respect to those which are local to an interaction.

Previous studies~\cite{AmelinoCamelia:2011cv,Loret:2014uia,Mignemi:2016ilu} of photon time delays in the context of deformations of SR have used a model based on worldlines of free particles that propagate in a noncommutative spacetime, which is a simple way to implement nontrivial translations which are compatible with the deformed Poincaré algebra. They have obtained different conclusions about the existence of photon time delays, apparently depending on the type of noncommutative spacetime under analysis. In the present paper we will consider the propagation in generic noncommutative spacetimes and establish the conditions for the absence of a photon time delay. As we will see, this may happen independently of the spacetime structure of the DSR theory, so that even first-order planckian corrections could be compatible with the above mentioned experimental results in a deformation of SR. As we will argue below, this fact opens up the possibility 
of a phenomenologically consistent deformation of SR with a new mass scale many orders of magnitude below the Planck mass.

The structure of the paper is as follows: in Sec.~\ref{sec:model} we will define the model and derive the expression for photon time delays. Then, in Sec.~\ref{sec:discussion} we will see how previous studies in $\kappa$-Minkowski and Snyder spacetimes~\cite{AmelinoCamelia:2011cv,Loret:2014uia,Mignemi:2016ilu} are particular cases of this model and give specific conditions for the absence of energy-dependent photon time delays. Finally, we will conclude in Sec.~\ref{sec:conclusions}.

\section{A general model for photon time delays in noncommutative spacetimes}
\label{sec:model}

The study of time delay effects in the propagation of particles in a modification of SR needs the specification of a spacetime.
Effective field theories that violate Lorentz invariance~\cite{Kostelecky:2008ts,*Long:2014swa,*Kostelecky:2016pyx,*Kostelecky:2016kkn} consider the classical, commutative, spacetime of SR. As argued above, this is no longer possible in the context of theories that modify SR but maintain a relativity principle, since they must incorporate nontrivial translations in spacetime. This is why noncommutativity is usually regarded as an appropriate feature of the spacetime formulation of DSR. While this ingredient is not yet worked out in full generality,\footnote{DSR theories are naturally formulated in momentum space~\cite{Amelino-Camelia2002,*Magueijo2002,*AmelinoCamelia:2002gv}.} there are specific examples of noncommutative spacetimes that have been explored and considered as benchmarks for the spacetime structure of a relativistic theory beyond SR. These include $\kappa$-Minkowski spacetime and the associated momentum space of the $\kappa$-Poincaré algebra, and also Snyder spacetime (both examples are reviewed in~\cite{KowalskiGlikman:2002jr}, and they were considered in relation with the calculation of photon time delays in Refs.~\cite{AmelinoCamelia:2011cv,Loret:2014uia} and~\cite{Mignemi:2016ilu}, respectively). In these cases, not only spacetime, but the whole phase space structure is modified with respect to the canonical phase space of SR.\footnote{In this paper we do not consider a possible role of the curvature of spacetime. A curvature of spacetime compatible with deformed relativistic theories has only recently begun to be explored in connection with Finsler geometries~\cite{Amelino-Camelia:2014rga,*Lobo:2016xzq} or a Hamiltonian formalism~\cite{Barcaroli:2015xda,*Barcaroli:2016yrl}.}

If $(x,p)$ are the spacetime and momentum coordinates of a canonical phase space,
\be
\{p_\mu,x^\nu\}=\delta_\mu^\nu,\quad \{x^\mu,x^\nu\}=\{p_\mu,p_\nu\}=0,
\ee
one can construct a nontrivial spacetime ($\tilde{x}$) by considering a linear combination of the $x$ coordinates with coefficients depending on the momentum variables $p$ and a new scale $M$,\footnote{For recent works using this construction see Ref.~\cite{Meljanac:2016jwk,*Loret:2016jrg}.} which must be necessarily introduced from dimensional arguments:
\be
\tilde{x}^\mu \,=\, x^\nu \, \varphi^\mu_\nu(p/M)\,.
\label{eq:NCdef}
\ee  
One can refer to this new space as ``noncommutative'' in the sense that the Poisson bracket of two of these coordinates is not zero for a generic choice of coefficients $\varphi^\mu_\nu$.  If we calculate the Poisson bracket of two spacetime coordinates we find
\be
\lbrace \tilde{x}^\mu, \tilde{x}^\nu\rbrace \,=\, -x^\sigma \varphi^\mu_\rho \frac{\partial\varphi^\nu_\sigma}{\partial p_\rho} + x^\rho \varphi^\nu_\sigma \frac{\partial\varphi^\mu_\rho}{\partial p_\sigma}\,.
\ee 

The worldlines of a particle $x^\mu(\tau)$, with $\tau$ an arbitrary parameter that flows along the worldline, can be obtained by applying the variational principle to the action
\be
\mathcal{S}=\int d\tau \left[\dot{x}^\mu p_\mu - N(\tau) (C(p)-m^2)\right]\,.
\label{eq:action}
\ee
This is the usual action of a free particle in SR with the substitution of $p^2$ by the function $C(p)$ which will define the modified dispersion relation, as one can see by taking the derivative of $\mathcal{S}$ with respect to the Lagrange multiplier $N(\tau)$:
\be
-\frac{\delta S}{\delta N(\tau)}=C(p)-m^2=0\,,
\ee
where the derivative has been equaled to zero because the action must be stationary. When this condition is applied to the derivative of $\mathcal{S}$ with respect to $x^\mu$ one gets that $p_\mu$ is constant along the worldline, and from the derivative with respect to $p_\mu$, one gets
\be
\dot{x}^\mu=N(\tau)\frac{\partial C}{\partial p_\mu}\,.
\label{eq:slope}
\ee 
The coefficient $N(\tau)$ was introduced in Eq.~\eqref{eq:action} to make the action invariant under reparameterizations $\tau\to \tau'=f(\tau)$. Once we get Eq.~\eqref{eq:slope}, we can choose $N(\tau)$ to take any particular value; the choice $N(\tau)=1$ leads to $\dot{x}^\mu$ constant along the worldline.

We can now obtain the worldline in the noncommutative spacetime:
\be
\tilde{x}^\mu(\tau)=\dot{\tilde{x}}^\mu \tau + \tilde{x}^\mu(0)=\varphi^\mu_\nu(p)\frac{\partial C}{\partial p_\nu} \tau+\tilde{x}^\mu(0),
\label{eq:WL}
\ee
where we just used Eq.~\eqref{eq:NCdef} and took into account that $p_\mu$ is constant along the worldline.

We could also have obtained the previous result from the Poisson bracket of the function $C(p)$ defining the dispersion relation, $C(p)=m^2$, with $\tilde{x}^\mu$:\footnote{This amounts to consider $C(p)$ as the generator of the evolution in $\tau$, which is the natural interpretation of Eq.~\eqref{eq:action}, in which $C(p)$ appears multiplying the coefficient $N(\tau)$ which implements the invariance of the action under reparameterizations.}
\be
\dot{\tilde{x}}^\mu=\{C,\tilde{x}^\mu\}=\frac{\partial C}{\partial p_\nu}\{p_\nu,\varphi^\mu_\rho(p) x^\rho\}=\varphi^\mu_\nu(p)\frac{\partial C}{\partial p_\nu}\,.
\ee
We can then define the velocity vector of the particle\footnote{Note that the velocity is independent of the choice of the parameter $\tau$.}
\be
\tilde{v}^i=\frac{\dot{\tilde{x}}^i}{\dot{\tilde{x}}^0}=\frac{\varphi^i_\nu (\partial C(p)/ \partial p_\nu)}{\varphi^0_\nu(\partial C(p)/\partial p_\nu)}\,.
\label{eq:v3d}
\ee

\subsection{Determination of time delays}

Our model considers the worldline of a free particle (a photon) that has its origin at a source (the emission point) and its end at a detector (the detection point). Since there is only one vector (the momentum $\vec{p}$ of the particle), the problem can be treated in $1+1$ dimensions without any loss of generality, so we will speak of its energy $E$ and its momentum $p\equiv |\vec{p}|$.\footnote{Although we use the same notation ($p$) for the four-momentum in 3+1 and for the momentum in 1+1, one can easily identify from the context which one is involved in the different equations.} We assume that low-energy photons (in the limit $p/M \to 0$) behave as in SR: that is, their time of travel equals (in natural units, $c=1$) the distance between source and detector. In fact, we will define this distance $L$ from the emission at the source and absorption at the detector of low-energy photons, for which the functions $\varphi^\mu_\nu \to \delta^\mu_\nu$, so that they observe a commutative spacetime $(\tilde{x}^\mu\to x^\mu)$.

The time delay of a high-energy photon of momentum $p$ at the detector with respect to a low-energy photon emitted ``simultaneously at the same point'' has two different sources: the modified dispersion relation $C(p)=0$, which defines the slope of the worldline, and the definition of simultaneity and spatial locality, which in a noncommutative spacetime, defined by the functions $\varphi(E,p)$ of Eq.~\eqref{eq:NCdef}, are relative concepts for observers whose spacetime origin do not coincide. In this case, there are two observers that must be brought into play: observer $A$, which is at the source, and observer $B$, which is at the detector. We define the origin of observer $A$ as the time and location of both a high and a low energy photon (this is the definition of a simultaneous emission at the source) and define the spacetime origin of observer $B$ to coincide with the detection of the low-energy photon. 

One can obtain the translation relating the noncommutative spacetime coordinates of both observers from the trivial translations relating the commutative coordinates, $x^B=x^A-L$, $t^B=t^A-L$ and Eq.~\eqref{eq:NCdef}:
\begin{align}
\tilde{t}^B& =\varphi^0_0 t^B+\varphi^0_1 x^B=\tilde{t}^A-L(\varphi^0_0 + \varphi^0_1) \label{traslaciont} \,,\\
\tilde{x}^B& =\varphi^1_0 t^B+\varphi^1_1 x^B=\tilde{x}^A-L(\varphi^1_0 + \varphi^1_1) \label{traslacionx} \,.
\end{align}

The worldline of the high energy particle for observer $A$ is
\be
\tilde{x}^A=\tilde{v}\,\tilde{t}^A\,,
\label{eq:AWL}
\ee
since $\tilde{x}^A=0, \tilde{t}^A=0$ corresponds to the initial point of the worldline, and $\tilde{v}$ is obtained particularizing Eq.~\eqref{eq:v3d} to $1+1$ dimensions:
\be
\tilde{v}=\frac{\varphi^1_0 (\partial C/\partial E)-\varphi^1_1(\partial C/\partial p)}{\varphi^0_0 (\partial C/\partial E)-\varphi^0_1(\partial C/\partial p)}\,,
\ee
where the minus signs appear due to the fact that $p_1=-p^1=-p$, and so $\partial C/\partial p_1=-\partial C/\partial p$. 
We can now obtain the worldline for observer $B$ by applying Eqs.~\eqref{traslaciont} and~\eqref{traslacionx} to Eq.~\eqref{eq:AWL}:
\be
\tilde{x}^B=\tilde{x}^A-L(\varphi^1_0 + \varphi^1_1)=\tilde{v}\,[\tilde{t}^B+L(\varphi^0_0 + \varphi^0_1)]-L(\varphi^1_0 + \varphi^1_1)\,.
\label{eq:BWL}
\ee
The end of the worldline for observer $B$ happens at $\tilde{x}^B=0$.\footnote{We are assuming that the detector is at rest. Then the spatial location for the detection of the high energy particle and the low energy photon coincide.} We can then obtain the value of $\tilde{t}^B$ at that point from Eq.~\eqref{eq:BWL}; this will give us the time-delay $\tilde{T}\equiv \tilde{t}^B(\tilde{x}^B=0)$ with respect to the arrival of the low-energy photon (that took place at $\tilde{t}^B=0$):
\be
\tilde{T}=\tilde{v}^{-1} \,L(\varphi^1_0 + \varphi^1_1)-L(\varphi^0_0 + \varphi^0_1)=L\left[(\varphi^1_0 + \varphi^1_1)\frac{\varphi^0_0 (\partial C/\partial E)-\varphi^0_1(\partial C/\partial p)}{\varphi^1_0 (\partial C/\partial E)-\varphi^1_1(\partial C/\partial p)}-(\varphi^0_0 + \varphi^0_1)\right]\,.
\label{eq:time-delay}
\ee 
Note that this equation is in fact valid for the time delay of any high-energy particle, independently of whether it is a photon or not. In the case of a particle of nonzero mass, Eq.~\eqref{eq:time-delay} includes both the usual special relativistic time-delay of a massive particle and the time-delay induced by new physics with respect the arrival of a low-energy photon emitted simultaneously at the source. 

\subsection{Momenta as generators of translations in spacetime}

The functions $\varphi^\mu_\nu$ that were introduced in Eq.~\eqref{eq:NCdef} are in correspondence with the Poisson brackets of energy and momentum with the noncommutative spacetime coordinates:
\be
\{E,\tilde{t}\}=\{E,\varphi^0_\nu x^\nu\}=\varphi^0_0\,, \quad \quad \{E,\tilde{x}\}=\varphi^1_0\,, \quad \quad
\{p,\tilde{t}\}=-\varphi^0_1\,, \quad  \quad \{p,\tilde{x}\}=-\varphi^1_1\,,
\ee
where again the minus signs appear due to $p_1=-p^1=-p$.

The transformations Eqs.~\eqref{traslaciont} and~\eqref{traslacionx} between observers $A$ and $B$ can then be written in the form
\begin{align}
\tilde{t}^B&=\tilde{t}^A-L\{E,\tilde{t}\}+L\{p,\tilde{t}\} \,,\\
\tilde{x}^B&=\tilde{x}^A-L\{E,\tilde{x}\}+L\{p,\tilde{x}\} \,.
\label{eq:transl}
\end{align}
These transformations correspond, then, to translations in the noncommutative spacetime which are generated by the energy and the momentum, even if the $(\tilde{x}^\mu,p_\nu
)$ phase space is noncanonical. This was in fact the approach taken in the analyses of Refs.~\cite{AmelinoCamelia:2011cv,Loret:2014uia} and~\cite{Mignemi:2016ilu}.

The formula for the time-delay Eq.~\eqref{eq:time-delay} can then be written in terms of Poisson brackets in the following form:
\begin{equation}
\tilde{T}=\left(L\{E,\tilde{x}\}-L\{p,\tilde{x}\}\right) \cdot \left(\frac{(\partial C/\partial E)\{E,\tilde{t}\}+(\partial C/\partial p)\{p,\tilde{t}\}}{(\partial C/\partial E)\{E,\tilde{x}\}+(\partial C/\partial p)\{p,\tilde{x}\}}\right)-L\{E,\tilde{t}\}+L\{p,\tilde{t}\}\,.
\label{eq:genTD}
\end{equation}

In a commutative spacetime, we have $\{E,t\}=1$, $\{E,x\}=0$, $\{p,t\}=0$, $\{p,x\}=-1$, and then Eq.~\eqref{eq:genTD} gives
\begin{equation}
T=-L\left(1+\frac{\partial C/\partial E}{\partial C/\partial p}\right)\,.
\label{eq:canTD}
\end{equation}
This expression contains of course the particular case of SR, $T=-L(1-E/p)$, which is zero for photons. 

\section{Discussion}
\label{sec:discussion}

We will now apply the previous results to several examples of deformations of the dispersion relation and of the Heisenberg algebra which have been considered in the literature. To this end, it is convenient to consider the approximation to the formula for the time delay, Eq.~\eqref{eq:genTD}, where one keeps the leading terms in these deformations. 
This can be done by writing the Poisson brackets as their canonical value plus an infinitesimal deformation of order $\epsilon$, $\{E,\tilde{t}\}=1+(\{E,\tilde{t}\}-1)=1+\mathcal{O}(\epsilon)$, $\{p,\tilde{x}\}=1+(\{p,\tilde{x}\}-1)=1+\mathcal{O}(\epsilon)$, $\{E,\tilde{x}\}=\mathcal{O}(\epsilon)$, $\{p,\tilde{t}\}=\mathcal{O}(\epsilon)$, writing also that $(\partial C/\partial E)/(\partial C/\partial p)=-E/p+\mathcal{O}(\epsilon)$, and expanding Eq.~\eqref{eq:genTD} neglecting $\mathcal{O}(\epsilon^2)$ terms. The result is:
\begin{equation}
\frac{\tilde{T}}{L} \approx -\left(1-\frac{E}{p}\right)-\left(\frac{\partial C/\partial E}{\partial C/\partial p}+\frac{E}{p}\right)-
\left(1-\frac{E}{p}\right)\left(\{E,\tilde{t}\}-1\right) + \left(1-\frac{E}{p}\right)\frac{E}{p} \,\{E,\tilde{x}\}\,.
\label{eq:genTDaprox}
\end{equation}
The first contribution is the time delay in SR, the second one is the correction due to the deformation of the dispersion relation, and the remaining last two contributions are due to the deformations in the Heisenberg algebra involving the energy variable $E$. The contributions due to deformations involving the momentum variable $p$ cancel out in the calculation of the time delay.

\subsection{A first example: photon time delays in $\kappa$-Minkowski spacetime}

$\kappa$-Minkowski spacetime has been considered as a model of a quantum spacetime that could emerge in a quantum theory of gravity~\cite{AmelinoCamelia:2008qg}. It is defined by the following commutation relations for the spatial ($\tilde{x}^i$) and time ($\tilde{x}^0\equiv \tilde{t}$) coordinates:
\begin{equation}
[\tilde{x}^0,\tilde{x}^i]=-\frac{i}{\kappa}\tilde{x}^i \,, \quad \quad [\tilde{x}^i,\tilde{x}^j]=0\,.
\label{eq:kM}
\end{equation}
With the identification of $\kappa\equiv M$, this corresponds to the Poisson bracket in $(1+1)$-dimensional spacetime
\be
\{\tilde{t},\tilde{x}\}=-M^{-1}\tilde{x} \,.
\ee

It is well known that $\kappa$-Minkowski is the noncommutative spacetime that is underneath the deformation of the Lorentz algebra known as $\kappa$-Poincaré~\cite{KowalskiGlikman:2002jr}. This fact was used in Ref.~\cite{AmelinoCamelia:2011cv} to try to answer the question of the momentum dependence on the speed of massless particles in this model of quantum spacetime. To this end, they worked in the, so-called, ``bicrossproduct basis'' of $\kappa$-Poincaré, where the $\kappa$-Minkowski deformed Casimir, at leading order in $\kappa^{-1}$, is
\begin{equation}
C(p)=p_0^2-\vec{p}^2-\frac{1}{M} p_0 \vec{p}^2\equiv m^2\,,
\label{eq:bicrossCasimir}
\end{equation}
and from Ref.~\cite{AmelinoCamelia:2011cv} we read the Poisson brackets of $\kappa$-Poincaré in the bicrossproduct basis in $1+1$ dimensions:
\begin{equation}
\{E,\tilde{t}\}=1 \,,\quad \quad \{E,\tilde{x}\}=0\,, \quad \quad \{p,\tilde{t}\}=-\frac{p}{M}\,, \quad \quad \{p,\tilde{x}\}=-1\,.
\end{equation}
When one uses the explicit form of the Casimir in the bicrossproduct basis [Eq.~\eqref{eq:bicrossCasimir}], one finds
\begin{equation}
\frac{\partial C/\partial E}{\partial C/\partial p}+\frac{E}{p}=\frac{1}{M}\left(\frac{E^2}{p}+\frac{p}{2}\right),
\label{eq:bicrossdispterm}
\end{equation}
and then Eq.~\eqref{eq:genTDaprox} gives
\begin{equation}
\frac{\tilde{T}[\text{bicross}]}{L}=-\left(1-\frac{E}{p}\right)-\frac{1}{M}\left(\frac{E^2}{p}+\frac{p}{2}\right)\,.
\label{eq:bicrossTD2}
\end{equation}
For photons, $m=0$ in Eq.~\eqref{eq:bicrossCasimir}, $E=p\,(1+p/2M)$ at first order, so that the time delay between a high-energy and a low-energy photon emitted simultaneously by the same source is $Lp/M$. This was the result obtained in Refs.~\cite{AmelinoCamelia:2011cv,Loret:2014uia}, which apparently concluded that propagation in $\kappa$-Minkowski spacetime implies an energy-dependent photon time delay.

However, we want to point out that the previous conclusion is not an inevitable consequence of the quantum nature of spacetime modeled by the $\kappa$-Minkowski commutation relations, but rather the result of the choice of a particular basis of the $\kappa$-Poincaré deformation algebra, which defines the deformed Casimir, that is, the particle dispersion relation. 
Since the $\kappa$-Poincaré algebra is a quantum nonlinear algebra, one can choose different bases related by arbitrary nonlinear transformations of momenta, all of them being algebraically equivalent. As Ref.~\cite{KowalskiGlikman:2002jr} showed, they may be chosen as energy-momentum sectors of different DSR theories, but all of them share the same noncommutative spacetime structure, which is given by the $\kappa$-Minkowski spacetime of Eq.~\eqref{eq:kM}.

A particular choice of basis in $\kappa$-Poincaré is the so-called ``classical basis'', for which the Casimir is the same as the one of special relativity, that is,
\begin{equation}
C(p)=p_0^2-\vec{p}^2\,,
\label{eq:classCasimir}
\end{equation}
and for which the Poisson brackets in $1+1$ dimensions are, at leading order~\cite{KowalskiGlikman:2002jr},
\begin{equation}
\{E,\tilde{t}\}=1 \,,\quad \quad \{E,\tilde{x}\}=-\frac{p}{M} \,,\quad \quad \{p,\tilde{t}\}=0 \,,\quad \quad \{p,\tilde{x}\}=-\left(1+\frac{E}{M}\right)\,.
\label{eq:PP}
\end{equation}
Eq.~\eqref{eq:genTDaprox} gives in this case
\begin{equation}
\frac{\tilde{T}[\text{class}]}{L}=-\left(1-\frac{E}{p}\right)\left(1+\frac{E}{M}\right)\,.
\label{eq:classTD}
\end{equation}
Therefore, for massless particles ($E=p$), there is not any time delay in the classical basis, even though the spacetime is noncommutative. 

Another basis proposed in Ref.~\cite{KowalskiGlikman:2002jr} is the Magueijo-Smolin basis. The modified Casimir at first order in this basis is 
\begin{equation}
C(p)=p_0^2-\vec{p}^2+\frac{1}{M}p_0^3-\frac{1}{M}p_0\vec{p}^2\,,
\label{eq:MG-SCasimir}
\end{equation}
and the Poisson brackets in $1+1$ dimensions are, at leading order, 
\begin{equation}
\{E,\tilde{t}\}=\left(1-\frac{2E}{M}\right)\,, \quad \quad \{E,\tilde{x}\}=-\frac{p}{M}\,, \quad \quad \{p,\tilde{t}\}=-\frac{p}{M}\,, \quad \quad \{p,\tilde{x}\}=-1\,.
\label{eq:PPM-S}
\end{equation}
From the expression of the Casimir [Eq.~\eqref{eq:MG-SCasimir}] one has
\begin{equation}
\frac{\partial C/\partial E}{\partial C/\partial p}+\frac{E}{p}=\left(1-\frac{E}{p}\right)\frac{E+p}{2M}\,,
\ee
and then   
\begin{equation}
\frac{\tilde{T}[\text{M-S}]}{L}=-\left(1-\frac{E}{p}\right)-\left(1-\frac{E}{p}\right)\frac{E+p}{2M}+\left(1-\frac{E}{p}\right)\frac{2E}{M}-\left(1-\frac{E}{p}\right)\frac{E}{M}=-\left(1-\frac{E}{p}\right)\left[1-\frac{E-p}{2M}\right]\,.
\label{eq:M-STD2}
\end{equation}
As in the classical basis, we see that for massless particles ($E=p$) there is not any time delay.

\subsection{Another example of noncommutativity: Snyder spacetime}

The first attempt to go beyond the continuum classical spacetime of SR maintaining Lorentz invariance was carried out by Snyder~\cite{Snyder:1946qz}. He proposed a noncommutative spacetime
\be
[\tilde{x}_\mu,\tilde{x}_\nu]=\frac{i}{M^2} J_{\mu\nu}\,,
\ee
with $J_{\mu\nu}$ the generators of the Lorentz algebra. 

As in the case of $\kappa$-Minkowski, there are different realizations in phase space. In the representation originally proposed by Snyder, the Poisson brackets in $1+1$ dimensions are
\begin{equation}
\{E,\tilde{t}\}=\left(1+\frac{E^2}{M^2}\right)\,, \quad \quad \{E,\tilde{x}\}=\frac{Ep}{M^2}\,, \quad \quad \{p,\tilde{t}\}=\frac{Ep}{M^2}\,, \quad \quad \{p,\tilde{x}\}=-\left(1-\frac{p^2}{M^2}\right),
\label{eq:PPSny}
\end{equation}
while in the representation of Maggiore~\cite{Maggiore:1993kv} at leading order:
\begin{equation} 
\{E,\tilde{t}\}=1+\frac{E^2-p^2}{2M^2}\,,\quad \quad \{E,\tilde{x}\}=0\,, \quad \quad \{p,\tilde{t}\}=0\,, \quad \quad \{p,\tilde{x}\}=-1-\frac{E^2-p^2}{2M^2}\,.
\label{eq:PPMagg}
\end{equation}
In both representations $C(p)$ is a function of $E$, $p$ only through the combination $(E^2-p^2)$. Then,
\begin{equation}
\frac{\partial C/\partial E}{\partial C/\partial p}+\frac{E}{p}=0,
\end{equation}
and one has in Eq.~\eqref{eq:genTDaprox}
\begin{equation}
\frac{\tilde{T}[\text{Snyder}]}{L}=-\left(1-\frac{E}{p}\right)-\left(1-\frac{E}{p}\right)\frac{E^2}{M^2}+\left(1-\frac{E}{p}\right)\frac{E}{p}\frac{Ep}{M^2}=-\left(1-\frac{E}{p}\right),
\label{eq:TD-Snyder}
\end{equation}
for the representation of Snyder, and
\begin{equation}
\frac{\tilde{T}[\text{Maggiore}]}{L}=-\left(1-\frac{E}{p}\right)-\left(1-\frac{E}{p}\right)\frac{E^2-p^2}{2M^2}=-\left(1-\frac{E}{p}\right)\left[1+\frac{E^2-p^2}{2M^2}\right] ,
\label{eq:TD-Maggiore}
\end{equation}
for the representation of Maggiore.

The conclusion is that there is no time delay for photons $(E=p)$ in Snyder spacetime, a result obtained recently~\cite{Mignemi:2016ilu} by identifying the coordinates in a curved momentum space of constant curvature that gives a geometric implementation of Snyder spacetime. We have shown that the result is contained as a particular case of the general expression~\eqref{eq:genTD} for the time delay in a model with a noncommutative spacetime.

\subsection{Condition for the absence of a photon time delay}

Having examined the most prominent examples in the literature, let us see now what happens when one considers generic modified dispersion relations and Poisson brackets at first order. The most general isotropic form\footnote{Isotropy allows to reduce trivially the model in $3+1$ dimensions to a model in $1+1$ dimensions.} for the functions $\varphi^{\mu}_{\nu}$ at first order is
\begin{equation}
\varphi^0_0=1+\frac{\delta_1}{M}p_0 \,,\qquad \varphi^0_i=\frac{\delta_2}{M}p_i \,,\qquad
\varphi^i_0=\frac{\delta_3}{M}p^i \,,\qquad
\varphi^i_j=\delta^i_j \left(1+\frac{\delta_4}{M}p_0\right)+\frac{\delta_5}{M}\epsilon^i_{jk}p^k\,,
\end{equation}
so that the Poisson brackets in 1+1 dimensions are
\begin{equation}
\{E,\tilde{t}\}=1+\frac{\delta_1}{M}E \,,\quad \quad \{E,\tilde{x}\}=\frac{\delta_3}{M}p \,,\quad \quad \{p,\tilde{t}\}=\frac{\delta_2}{M}p \,,\quad \quad \{p,\tilde{x}\}=-\left(1+\frac{\delta_4}{M}E\right)\,.
\label{eq:PPfirstorder}
\end{equation}
Using the most general modified rotational invariant dispersion relation  
\begin{equation}
C(p)=p_0^2-\vec{p}^2+\frac{\alpha_1}{M}p_0^3+\frac{\alpha_2}{M}p_0\vec{p}^2=m^2\,,
\label{eq:MDRfirstorder}
\end{equation}
one has
\begin{equation}
\frac{\partial C/\partial E}{\partial C/\partial p}+\frac{E}{p}=-\frac{E}{p}\left[\frac{3}{2}\frac{(\alpha_1+\alpha_2)}{M}E-\frac{\alpha_2}{2M}\frac{(E^2-p^2)}{E}\right]\,,
\label{eq:1st-order}
\end{equation}
so that, substituting in Eq.~\eqref{eq:genTDaprox}, one finds
\begin{equation}
\frac{\tilde{T}}{L}=-\left(1-\frac{E}{p}\right)+\frac{E}{p}\left[\frac{3}{2}\frac{(\alpha_1+\alpha_2)}{M}E-\frac{\alpha_2}{2M}\frac{(E^2-p^2)}{E}\right]-\left(1-\frac{E}{p}\right)\frac{\delta_1 E}{M}+\left(1-\frac{E}{p}\right)\frac{\delta_3 E}{M}\,,
\label{eq:firstTD}
\end{equation}
and for the massless case, $E/p\approx 1-(\alpha_1+\alpha_2)E/2M,$ one obtains
\begin{equation}
\tilde{T}=L\left(\alpha_1+\alpha_2\right)\frac{E}{M}\,.
\label{eq:firstmasslessTD}
\end{equation}
One can check that this result is consistent with the results obtained for the different bases studied previously. Equations \eqref{eq:firstTD} and \eqref{eq:firstmasslessTD} are valid either with or without an implementation of a relativity principle. 

One sees that the photon time delay is independent of the values of the Poisson brackets, and it is zero when  
\begin{equation}
\alpha_1+\alpha_2=0\,,
\label{eq:firstnullTD}
\end{equation}
independently of the spacetime under consideration.

We can do the same computation in the case in which the dominant correction starts at second order (proportional to $1/M^2$):
\begin{equation}
\begin{split}
\varphi^0_0&=1+\frac{\delta_6}{M^2}p^2_0+\frac{\delta_7}{M^2}\vec{p}^2 \,,\qquad 
\varphi^0_i=\frac{\delta_8}{M^2}p_0p_i \,,\qquad
\varphi^i_0=\frac{\delta_9}{M^2}p_0p^i\,,\qquad \\
\varphi^i_j&=\delta^i_j \left(1+\frac{\delta_{10}}{M^2}p^2_0+\frac{\delta_{11}}{M^2}\vec{p}^2\right)+\frac{\delta_{12}}{M^2}p^ip_j+\frac{\delta_{13}}{M^2}p_0\epsilon^i_{jk}p^k\,.
\end{split}
\end{equation}
In this case the Poisson brackets in $1+1$ dimensions are
\begin{equation}
\begin{split}
\{E,\tilde{t}\}&=1+\frac{\delta_6}{M^2}E^2+\frac{\delta_7}{M^2}p^2 \,,\quad \quad \{E,\tilde{x}\}=\frac{\delta_9}{M^2}Ep \,,\\
\{p,\tilde{t}\}&=-\frac{\delta_8}{M^2}Ep \,,\quad \quad \{p,\tilde{x}\}=-\left(1+\frac{\delta_{10}}{M^2}E^2+\frac{\delta_{11}+\delta_{12}}{M^2}p^2\right)\,.
\end{split}
\label{eq:PPsecondorder}
\end{equation}
The modified dispersion relation at second order is
\be
C(p)=p_{0}^{2}-{\vec{p}}^{2}+\frac{\alpha_{3}}{M^2}p_{0}^{4}+\frac{\alpha_{4}}{M^2}p_{0}^{2}\vec{p}^{2}+\frac{\alpha_{5}}{M^2}(\vec{p}^{2})^2=m^{2} \,,
\ee
which gives
\begin{equation}
\frac{\partial C/\partial E}{\partial C/\partial p}+\frac{E}{p}=-\frac{E}{p}\left[\frac{2(\alpha_3+\alpha_4+\alpha_5)}{M^2}E^2-\frac{(\alpha_4+2\alpha_5)}{M^2}(E^2-p^2)\right]\,,
\label{eq:1st-order}
\end{equation}
and substituting in Eq.~\eqref{eq:genTDaprox}, one gets
\begin{equation}
\frac{\tilde{T}}{L}=-\left(1-\frac{E}{p}\right)+\frac{E}{p}\left[\frac{2(\alpha_3+\alpha_4+\alpha_5)}{M^2}E^2-\frac{(\alpha_4+2\alpha_5)}{M^2}(E^2-p^2)\right]-\left(1-\frac{E}{p}\right)\left(\frac{\delta_6 E^2}{M^2}+\frac{\delta_7 p^2}{M^2}\right)+\left(1-\frac{E}{p}\right)\frac{\delta_9 E^2}{M^2}.
\label{eq:secondTD}
\end{equation}
For the massless case, $E/p\approx 1-(\alpha_3+\alpha_4+\alpha_5)E^2/2M^2$, one obtains
\begin{equation}
\tilde{T}=L\frac{3}{2}\left(\alpha_3+\alpha_4+\alpha_5\right)\frac{E^2}{M^2}\,.
\label{eq:secondmasslessTD}
\end{equation}
Once again, the value of the photon time delay is independent of the considered spacetime, and it is zero if
\begin{equation}
\alpha_3+\alpha_4+\alpha_5=0\,.
\label{eq:secondnullTD}
\end{equation}
This is indeed the condition that is satisfied in Snyder spacetime, when $C(p)$ is just a function of $(E^2-p^2)$.

\subsection{Interpretation of the results for time delays}

In all cases considered previously it can be seen that the time delay is proportional to 
\begin{equation}
L\left[(1+(\partial C/\partial E)/(\partial C/\partial p)\right]\,,
\label{factor}
\end{equation}
i.e., to $(L/v-L)$, where $v$ is the velocity of propagation of the high energy particle in the commutative spacetime,
\be
v=-\frac{\partial C/\partial p}{\partial C/\partial E}\,.
\label{velocidadC}
\ee 
This result can in fact be obtained from the general expression~\eqref{eq:time-delay}:
\be
\tilde{T}=L\left[(\varphi^1_0+\varphi^1_1)\frac{\varphi^0_0+\varphi^0_1 v}{\varphi^1_0+\varphi^1_1 v}-(\varphi^0_0+\varphi^0_1)\right]=\frac{L(\varphi^0_0\varphi^1_1-\varphi^1_0\varphi^0_1)}{\varphi^1_0+\varphi^1_1 v}(1-v)=\frac{\varphi^0_0\varphi^1_1-\varphi^1_0\varphi^0_1}{\varphi^1_1+\varphi^1_0/v}L\left(\frac{1}{v}-1\right).
\ee
Then we see that the time delay is just the naive result taking into account the (possible) energy dependence of the velocity of propagation, multiplied by a factor which involves the dependence on the details of the noncommutative spacetime. 

In the case of photons (for which there is no time delay in SR), and in an expansion in powers of $(1/M)$, the time delay will be proportional to $(1/M)$ [or $(1/M^2)$ if corrections start at second order]. But the factor given by expression~\eqref{factor} is already of order $(1/M)$ [or $(1/M^2)$ for a second order correction]. Then the dominant contribution to the time delay will be independent of the details of the noncommutative spacetime and will be determined just by the modification of the dispersion relation $C(p)$. This is the reason why the coefficients $\delta_i$ which parametrize the general form of the dominant term in an expansion for the modification of the Heisenberg algebra cancel in the final expression for the dominant contribution to photon time delays.    

The condition for the absence of time delays is, therefore, $v=1$. But, taking differentials in the photon dispersion relation, $C(E,p)=0$, one gets
\be
-\left.\frac{\partial C/\partial p}{\partial C/\partial E}\right|_{C(E,p)=0} \,=\, \frac{d E(p)}{d p}\,,
\ee
where $E(p)$ is the solution of $C(E, p)=0$. Then the absence of time delay requires that for photons, $E(p)=p$. 
The possibility to have modified dispersion relations that reduce to $E(p)=p$ for photons was already noted in Ref.~\cite{Hossenfelder:2005ed}. Quite generally,
if one considers an expansion in powers of $(1/M)$
\begin{equation} 
C(E,p) \,=\, E^2-p^2+\sum_{n}\frac{1}{M^n}C^{(n)}(E,p)\,,
\end{equation}   	
then the conditions that the dispersion relation has to satisfy so that photons do not show a time delay are
\begin{equation}
\left. C^{(n)}(E,p) \right|_{p=E} =0\,.
\end{equation}
One has $\alpha_1+\alpha_2=0$ for $n=1$, $\alpha_3+\alpha_4+\alpha_5=0$ for $n=2$ and similar expressions for higher order terms. This is consistent with the result obtained by a general calculation of time delays for a first (or a second) order correction to SR. 

In the case of a massive high energy particle one would have a time delay in SR due to the energy dependence of the velocity of propagation of a massive particle that could be modified due to corrections to SR. Even if the dispersion relation were not modified (and, therefore, $v$ were the same as in SR), one would have a time delay for massive particles which would be the mass dependent time delay of SR, multiplied by a factor depending on the details of the noncommutativity proportional to the ratio $E/M$ [or $(E^2/M^2)$]. In order to have an observation sensitive to the noncommutativity one would require a measurement of the mass dependent time delay with a precision of order $(E/M)$ [or $(E^2/M^2)$].     

\section{Conclusions}
\label{sec:conclusions}

The main result obtained in this work is that it is possible to go beyond SR without observable time delays, even for a deformation of SR (that is, in the presence of a relativity principle). We have identified that this is the case whenever the velocity of propagation of photons in the commutative spacetime is independent of the energy, which would be the naive conclusion forgetting about the effects of a nontrivial spacetime. Considering a model based on worldlines of particles in a noncommutative spacetime we have found several examples with no time delay for photons: $\kappa$-Minkowski in the classical basis, $\kappa$-Minkowski in the Magueijo-Smolin basis and Snyder spacetime in two different representations.

Although the standard studies of Lorentz violation assume a commutative spacetime, one could consider a noncommutative spacetime also in this case. The difference between a scenario with LIV and a relativistic theory is that in the latter case one has to keep a (deformed) Poincar\'e symmetry, which requires a consistency of the modified dispersion relation and the nontrivial implementation of translational symmetry which fixes the noncommutativity of spacetime. On the contrary, in the case of LIV one can choose independently the dispersion relation and the noncommutativity of spacetime. The absence of time delays is a property which depends exclusively on the dispersion relation and then applies indistinctly to both cases.

All the results of this work are based on a noncommutative spacetime as the appropriate model to calculate time delays induced by departures from SR. This is the most common framework that has been considered in previous analyses~\cite{AmelinoCamelia:2011cv,Loret:2014uia,Mignemi:2016ilu} in order to implement nontrivial translations in the context of deformations of SR, and it is the simplest way to introduce a relativity of locality for the emission and detection processes, which are local only for the observers at the source and at the detector, respectively. However, the framework introduced in Ref.~\cite{AmelinoCamelia:2011bm} considers a relativity of the locality associated to the interaction of particles as an alternative to a formulation based on spacetime noncommutativity. In this case one should consider the nonlocality of the interactions responsible for the emission and detection of particles as the appropriate model to study the possibility to have observable time delays. In a future work~\cite{Carmona:2016a} we will study time delays in this alternative framework to confirm the conclusions obtained in the present paper.

One could take the main outcome of this work as bad news from a phenomenological point of view since it means that if the possible departures from SR induced by quantum gravity are such that there is no time delay for photons 
the only phenomenological window to deformations of SR (if they do not include the possibility of photon birefringence) gets closed. Alternatively one can take the result as good news since it means that the strong constraints on the mass scale parametrizing the departures from SR obtained from the absence of observations of time delays for photons are not applicable with full generality. Such constraints would not necessarily imply that quantum gravity corrections start at least at second order of the Planck mass (which is still a possibility), but would be compatible with first-order modifications of SR which do not produce photon time delays.

Moreover, since all the constraints based on the use of effective field theory to go beyond SR~\cite{Kostelecky:2008ts,*Long:2014swa,*Kostelecky:2016pyx,*Kostelecky:2016kkn} apply only to an scenario without the presence of a relativity principle, the result obtained in this paper opens up the interesting possibility of a scale of deformation of special relativity much smaller than its simplest estimate (the Planck mass) without any phenomenological inconsistency.
The simplest estimate is based on naturalness but we already know that this argument fails in the case of the vacuum energy, and we have also hints that it also fails in the estimate for the mass of elementary scalars (the standard model (SM) Higgs particle seems to be much lighter than the scale limiting the domain of validity of the SM). Then it seems reasonable to explore the possibility of some mechanism generating a scale for the departures from SR much smaller than this simplest estimate. If this were the case then one would have to reconsider searches of possible signals of quantum gravity that may have been discarded based on the assumption that the scale for such signals is of the order of the Planck mass.

\section*{Acknowledgments}
This work is supported by the Spanish MINECO FPA2015-65745-P (MINECO/FEDER) and Spanish DGIID-DGA Grant No. 2015-E24/2. We acknowledge useful conversations with Niccolò Loret.

%\bibliography{LIV,Mypapers}
%%%%%%%%%%%%%%%%%%%%%%%%%%%%%%%%%%%%%%%%

%merlin.mbs apsrev4-1.bst 2010-07-25 4.21a (PWD, AO, DPC) hacked
%Control: key (0)
%Control: author (72) initials jnrlst
%Control: editor formatted (1) identically to author
%Control: production of article title (-1) disabled
%Control: page (0) single
%Control: year (1) truncated
%Control: production of eprint (0) enabled
%

%%%%%%%%%%%%%%%%%%%%

\end{document}